# Spin echo from erbium implanted silicon

Mark A. Hughes,[1,a)] Naitik A. Panjwani,[2,3] Matias Urdampilleta,[2,4] Kevin P. Homewood,[5,6] Ben Murdin,[5] and J. David Carey[5,7]

**AFFILIATIONS**

[1]Materials and Physics Research Group, School of Science, Engineering and Environment, University of Salford, Salford M5 4WT, United Kingdom
[2]University College London, London Centre for Nanotechnology, Gower Place, London WC1E 6BT, United Kingdom
[3]Berlin Joint EPR Lab, Fachbereich Physik, Freie Universität Berlin, D-14195 Berlin, Germany
[4]Institut Néel-CNRS-UJF-INPG, UPR2940, 25 rue des Martyrs, BP 166, 38042 Grenoble Cedex 9, France
[5]Advanced Technology Institute, Faculty of Engineering and Physical Sciences, University of Surrey, Guildford GU2 7XH, United Kingdom
[6]School of Materials Science and Engineering, Hubei University, Wuhan 430062, People's Republic of China
[7]Department of Electrical and Electronic Engineering, University of Surrey, Guildford GU2 7XH, United Kingdom

[a)]Author email m.a.hughes@salford.ac.uk

**ABSTRACT**

Erbium implanted silicon as a quantum technology platform has both telecommunications and integrated circuit processing compatibility. In Si implanted with Er to a concentration of $3 \times 10^{17}\,\text{cm}^{-3}$ and O to a concentration of $10^{20}\,\text{cm}^{-3}$, the electron spin coherence time, $T_2$, and the spin-lattice relaxation time, $T_1$, were measured to be 7.5 $\mu$s and $\sim$1 ms, respectively, at 5 K. The spin echo decay profile displayed strong modulation, which was consistent with the super-hyperfine interaction between $Er^{3+}$ and a spin bath of $^{29}Si$ nuclei. The calculated spectral diffusion time was similar to the measured $T_2$, which indicated that $T_2$ was limited by spectral diffusion due to $T_1$-induced flips of neighboring $Er^{3+}$ spins. The origin of the echo is an Er center surrounded by six O atoms with monoclinic $C_{1h}$ site symmetry.

The optical fiber telecommunications network makes telecom wavelength photons at 1.5 $\mu$m by far the best candidate for transferring quantum information over distance. $Er^{3+}$ intra-4f shell transitions can be optically addressed at telecommunications C band wavelengths, which would allow the transfer of quantum information over distance. The use of rare earth (RE) ions is well suited to overcome a paradox of quantum technology (QT) platform requirements: sufficient decoupling from the environment to avoid decoherence, but a strong enough interaction with the environment to allow addressing, readout, and gating. The advantage of RE ions arises as they possess a partially filled 4f shell that is shielded from the environment by the outer 5s and 5p shells, leading to extraordinarily coherence times of 6 h for optically detected nuclear spin[1] and 4.4 ms for optical transitions;[2] however, even with their atomic scale shielding, long lived entanglement between RE dopants in a solid matrix has been observed,[3,4] and entanglement between internal degrees of freedom of single RE ions, with a less than half filled 4f shell, can still exist up to thousands of kelvin, making this one of the most stable known entanglements.[5]

Architectures based on Si offer considerable promise for developing a practical and scalable pathway for quantum computer fabrication, and features can be patterned in Si on the scale required for many quantum device architectures. Ion implantation of Si is a well understood technology in integrated circuit (IC) fabrication, and commercial adoption of new technologies tends to favor those based on established fabrication platforms and techniques. Recently, increases in coherence times by several orders of magnitude have been demonstrated in donor impurities in silicon by using isotopically pure $^{28}Si$.[6] However, these donor impurities do not interact with light at telecommunications wavelengths, which is critical for many quantum communication schemes. A possible alternative is the T-center in Si, thought to be composed of two carbon atoms, which displays spin-selective bound exciton optical transitions at 1326 nm.[7] Given expected improvements in $T_2$ by using $^{28}Si$, optimizing processing for the appropriate Er-related centre[8] and reducing Er concentration, Er implanted Si is potentially the only known QT platform with telecommunications C band addressability, long $T_2$, and IC tooling compatibility. The spin state of a single Er ion implanted into a silicon single electron transistor has been optically addressed and electrically readout,[9] whereas the spin state of a single Er ion in $Y_2SiO_5$, coupled to a silicon nanophotonic cavity, can be read out optically with a single shot.[10] This demonstrates that Er implantation is compatible with single electron transistors, which are used in the readout of quantum dot qubits.[11] Here we report spin echo measurements of Er implanted Si.

TABLE I. The principal g values and tilt angles, τ, of Er doped Si determined by ESR and Zeeman measurements.

| Center | Symmetry | $g_x$ | $g_y$ | $g_z$ | τ | References |
|---|---|---|---|---|---|---|
| ESR | | | | | | |
| OEr-1 | Monoclinic $C_{1h}$ | 0.8 | 5.45 | 12.6 | 57.30° | 14 |
| OEr-1' | Monoclinic $C_{1h}$ | 0.8 | 5.45 | 12.55 | 56.90° | 14 |
| OEr-3 | Monoclinic $C_{1h}$ | 1.09 | 5.05 | 12.78 | 48.30° | 14 |
| OEr-4 | Trigonal $C_{3v}$ | 2.0 | 6.23 | 6.23 | 54.74° | 14 |
| OEr-2 | Trigonal $C_{3v}$ | 0.45 | 3.46 | 3.22 | 55.90° | 14 |
| OEr-2' | Trigonal $C_{3v}$ | 0.69 | 3.24 | 3.24 | 54.74° | 14 |
| Zeeman | | | | | | |
| Er-1 | Orthorhombic $C_{2v}$ | ∼0 | ∼0 | 18.4 | 45° | 19,21 |

A sample, with an Er concentration of $3 \times 10^{17}$ $cm^{-3}$ and an O concentration of $10^{20}$ $cm^{-3}$, was prepared by implanting Er and O ions into a ⟨100⟩ oriented 8000 ± 500 Ω cm Si wafer. The sample was then annealed at 750 °C for 2 min to recrystallize the amorphized region. O and Er ions were implanted at a range of energies to give a flat concentration profile down to a depth of around 1.5 μm, see the supplementary material, Fig. S1. The uncertainty in the Er dose, the accuracy of the implant simulation, and diffusion after annealing[12] contribute to an uncertainty in the Er concentration of ±10%. Isotope specific implantation was used so that only the zero nuclear spin $^{166}Er$ was implanted.

CW and pulsed ESR measurements were performed in a Bruker E580 ESR spectrometer. All ESR measurements were recorded with the magnetic field, $B_0$, parallel to the [001] direction of the wafer with an uncertainty of ±5°. For pulsed measurements, the Q factor was detuned to ∼100; the π pulse width and repetition time were and 32 ns, and 4.5 ms, respectively. Phase cycling was used in pulsed measurements, but this failed to remove an off resonance echo signal, which was also present in empty tube measurements. This off resonance echo signal was subtracted during analysis. All CW ESR measurements were made at 10 K, all pulsed measurements at 5 K, and the microwave frequency was 9.61 GHz.

When implanted into Si, Er exists in its usual 3+ oxidation state.[13] Oxygen was co-implanted to a concentration of $10^{20}$ $cm^{-3}$ and is required to generate narrow Er-related ESR[14] and photoluminescence (PL)[15] lines by the creation of various O coordinated Er (Er–O) centers. Previous measurements of the angular dependence of the Er-related ESR lines in Er implanted Si have identified a number of different Er–O ESR centers: three monoclinic centers labeled OEr-1, OEr-1', and OEr-3, and three trigonal centers labeled OEr-2, OEr-2', and OEr-4.[14,16–18] Also, Zeeman measurements of molecular beam epitaxy (MBE) grown Er doped Si have identified an orthorhombic Er–O center, labeled Er-1,[19] which, possibly because of a long $T_1$, is not ESR active.[20] We have not yet linked any of the ESR centers to the numerous PL lines observed in Er implanted Si; however, the Zeeman ground state of the Er-1 center can be readily populated by 1.5 μm laser radiation,[20] which should also be feasible in the ESR centers.

The principal g values of these centers are given in Table I. For all ESR centers, the $g_y$-axis is parallel to a [1 1 0] direction and the mutually perpendicular $g_x$- and $g_z$-axis lie in the (1 1 0) plane with the $g_x$-axis tilted away from [0 0 1] by an angle τ.

Figure 1(a) shows the measured and simulated CW ESR, and the field dependent echo. There are two main resonances at 867 and 934 G, respectively, with weaker resonances at 892 and 964 G. We simulated the ESR spectra of each center with their g-values and tilt angle using EASYSPIN;[22] the code can be found in the associated dataset. Only the OEr-3 center can explain the observed resonances. We simulated varying the rotation angle (angle between [001] and $B_0$ in the $(1\bar{1}0)$ plane) and found the closest match between simulated and observed ESR spectra with a rotation angle of 5.2°. The discrepancy between the simulated and measured spectra should be due to some small differences in the g-values and tilt angle of our $3 \times 10^{17}$ cm$^{-3}$ Er sample, and the $10^{19}$ cm$^{-3}$ Er sample in Ref. 14. All of the measured ESR resonances are visible in the echo spectrum, showing that the echo originates from the Er ESR centers. The Er-related ESR lines have been attributed to an $Er^{3+}$ center based on similar g-values to Er doped $Y_2O_3$, which has the same crystal structure as $Er_2O_3$, and EXAFS measurements of Er and O implanted Si, which found a similar Er–O bond length to that in $Er_2O_3$.[14,18] A spin echo signal was also present off resonance and was also present in empty tube measurements; we therefore treated the off resonance echo as a background and subtracted it, as illustrated in supplementary material Fig. S2.

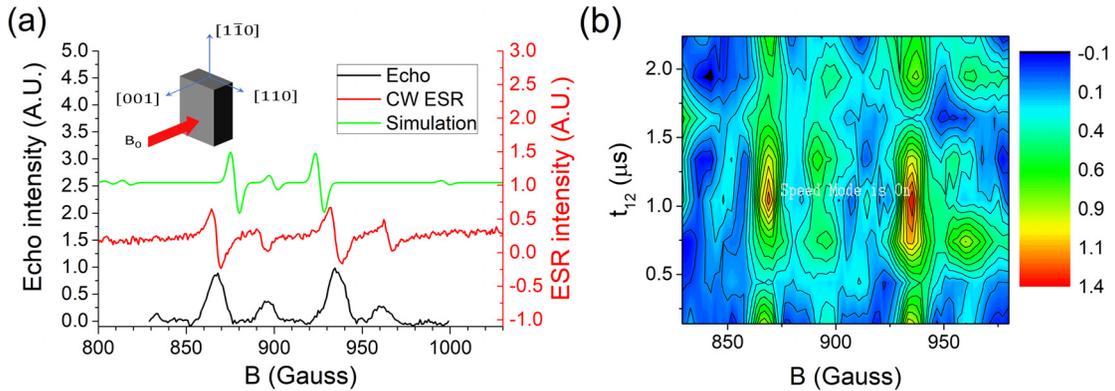

FIG. 1. (a) Simulated CW ESR of OEr-3 center, measured CW ESR, and field dependent echo ($t_{12}$ = 130 ns) signal. (b) Contour plot showing the echo intensity as a function of magnetic field at various $t_{12}$. All CW ESR measurements were made at 10 K, all echo measurements at 5 K, and the microwave frequency was 9.61 GHz.

Figure 1(b) shows the echo intensity as a function of $B_0$ for various $t_{12}$ between 0.14 and 2.24 μs. The on-resonance echo signal disappears below the detection limit and then reappears with increasing $t_{12}$, indicating the presence of strong electron spin echo envelope modulation (ESEEM).

In Fig. 2(a) we show the echo decay profile at a fixed $B_0$ of 867 G, corresponding to the OEr-3 center. The background echo signal was subtracted as shown in supplementary material Fig. S3 to give a recovered resonance echo decay. As can be seen from Fig. 1(a), a subtraction of the echo signal at 850 G from that at 867 G should leave only the echo component attributable to implanted Er. The echo decay in Fig. 2(a) displays strong superimposed oscillations from the ESEEM effect[23] caused by superhyperfine coupling with neighboring nuclear spins; similar oscillations were observed in $Er:CaWO_4$, but were significantly weaker than those seen here.[24] Since the 4f wavefunction is highly localized, the superhyperfine coupling between a RE and a neighboring nuclear spin is usually regarded as magnetic dipole–dipole only.[25,26] The ESEEM effect when using to a two-pulse echo sequence, $V_{2p}(t_{12})$, on an isolated $Er^{3+}$ ion (effective electron spin $S = \frac{1}{2}$) in proximity to a $^{29}Si$ nuclei (nuclear spin $I = \frac{1}{2}$) can be described as follows:[23]

$$V_{2p}(t_{12}) = 1 - \frac{k}{4}[2 - 2\cos(2\pi\nu_\alpha t_{12}) - 2\cos(2\pi\nu_\beta t_{12}) + \cos(2\pi(\nu_\alpha - \nu_\beta)t_{12}) + \cos(2\pi(\nu_\alpha + \nu_\beta)t_{12})], \quad (1)$$

where $k$ is the modulation index and $\nu_\alpha$ and $\nu_\beta$ are the $^{29}Si$ nuclear resonance frequencies for the two possible $Er^{3+}$ electron spin orientations ($S = \pm\frac{1}{2}$). The OEr-3 center, along with the other ESR centers, is thought to be O coordinated because they are only observed when O is co-implanted to a concentration $10^{20}$ cm$^{-3}$,[27] and EXAFS measurements of Er and O co-implanted Si showed that the average number of O atoms surrounding the Er atom was $5.1 \pm 0.5$, and the average Er–O separation was 2.26 Å.[28] The strong ESEEM modulation observed in Fig. 2(a) contains two frequencies: $\nu_1 = 0.70$ MHz and $\nu_2 = 1.66$ MHz, as shown in the FFT in Fig. 2(b); the nuclear spin Larmor frequency ($\nu_L$) is 0.76 MHz for $^{29}Si$. One possible origin of $\nu_1$ and $\nu_2$ is $^{29}Si$ at a single crystallographic position with $\nu_\alpha = \nu_1 \approx \nu_L$ and $\nu_\beta = \nu_2 \approx 2\nu_L$. This is unlikely because a large difference in $\nu_\alpha$ and $\nu_\beta$ indicates very strong coupling, which would be expected from nuclei in the first coordination sphere, which in this case is O. Also, the large modulation index is inconsistent with the 4.67% abundance of $^{29}Si$ and a single crystallographic position. Another possibility is $\nu_1$ and $\nu_2$ originating from two crystallographic positions, one with $\nu_\alpha \approx \nu_\beta \approx \nu_1 \approx \nu_L$ and the other with $\nu_\alpha \approx \nu_\beta \approx \nu_2 \approx 2\nu_L$. This is unlikely because $\nu_\alpha \approx \nu_\beta \approx \nu_L$ indicates weak coupling, which, for a single crystallographic position, is inconsistent with the large modulation index. Also, a crystallographic position with $\nu_\alpha \approx \nu_\beta \approx 2\nu_L$ indicates strong coupling, which can be discounted for the same reasons given for the first possible origin. Given the structure of the OEr-3 center, it is more likely that the observed ESEEM is due to weak coupling to many $^{29}Si$ nuclei, in which case each $^{29}Si$ nuclei has $\nu_\alpha \approx \nu_\beta \approx \nu_L$ and $k \ll 1$; in this situation, the overall decay profile is given by[29]

$$I(t_{12}) = I_0 e^{-\left(\frac{2t_{12}}{T_2}\right)^x}\left[1 - \frac{k}{4}[3 - 4\cos(2\pi\nu_L t_{12}) + \cos(4\pi\nu_L t_{12})]\right]. \quad (2)$$

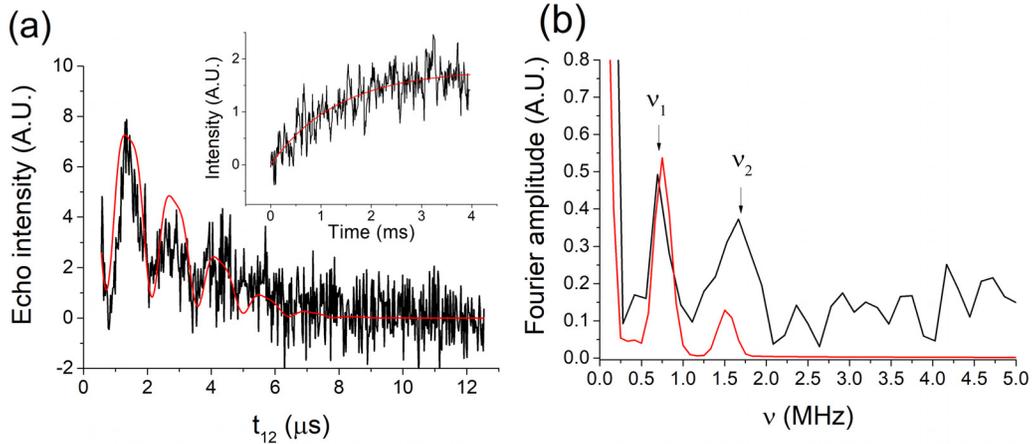

**FIG. 2.** (a) Spin echo decay profile at a $B_0$ of 867 G, with the off resonance decay profile subtracted (black), fitted to Eq. (2) (red), with x fixed to 2 and fit parameters of $T_2 = 7.5 \pm 3$ μs, $k = 0.46 \pm 0.1$, $\nu_L = 0.77 \pm 0.05$ MHz. Inset shows the saturation recovery at 867 G, with the off resonance saturation recovery subtracted, (black) fitted with a single exponential fit (red) to give $T_1$ of $0.98 \pm 0.2$ ms. (b) FFT of measured (black) and fitted (red) decay profiles; the position of the two main peaks in the FFT of the measured decay reveals two frequencies: $\nu_1 = 0.70$ MHz and $\nu_2 = 1.66$ MHz. All measurements were at 5 K, and the microwave frequency was 9.61 GHz.

The $I_0 e^{-\left(\frac{2t_{12}}{T_2}\right)^x}$ term is the empirical Mims equation,[30] which describes the echo decay in the absence of nuclear coupling where $T_2$ is the spin coherence time and $x$ is an exponential stretch factor that is determined by spin dynamics.

In Fig. 2(a), fitting to Eq. (2) was consistent with the observed echo decay profile and the FFT of the measured and fitted decay in Fig. 2(b) were consistent with each other and indicate that $\nu_2$ is the $2\nu_L$ component of Eq. (2). The FFT peaks are broadened by the relatively short sampling time; the larger relative Fourier amplitude of $\nu_2$ in the FFT of the measured decay is a result of the low signal to noise ratio (SNR) of our data and our choice of a Hanning window, which has relatively poor amplitude accuracy, but good spectral resolution.[31] When we used a flat top window, the amplitudes were much closer to the fitted FFT, but the spectral resolution was inadequate. Fitting yielded $k = 0.45$, which we expect to be composed of the overall modulation effect of many crystallographic positions with $k \ll 1$. The fit also yielded a $T_2$ of $7.5 \pm 3$ $\mu$s, which compares to $\sim 5$ $\mu$s at 5 K ($\sim 50$ $\mu$s at 2.5 K) for $\sim 10^{16}$ cm$^{-3}$ Er doped CaWO$_4$,[24] and 1.6 $\mu$s at 1.9 K for $\sim 2 \times 10^{18}$ cm$^{-3}$ Er doped Y$_2$SiO$_5$.[32] This is a promising comparison given the difficulty of removing defects after implantation that could lead to decoherence. Further optimization of the recrystallization process, reductions in Er concentration and isotopic purification of the Si may lead to coherence times applicable to quantum communication and computation.

The saturation recovery profile shown in the inset of Fig. 2(a) gives a spin relaxation time, $T_1$, at 5 K, of $0.98 \pm 0.2$ ms; the background subtraction is shown in supplementary material Fig. S4. Spectral diffusion often limits $T_2$ and can be caused by various electron[33] and nuclear[34] spin flip-flop process. The spectral diffusion time ($T_{SD}$) due to $T_1$-induced flips of neighboring Er$^{3+}$ spins can be calculated from the Er$^{3+}$ concentration in cm$^{-3}$ ([Er]), $T_1$, and effective g-factor ($g_e$), in this case at 867 G, using the following equation:[30,35]

$$T_{SD} = \sqrt{\frac{18\sqrt{3}}{4\pi^2} \frac{\hbar}{(g_e \beta_e)^2} \frac{T_1}{[Er]}}. \quad (3)$$

This yields a $T_{SD}$ of 7.11 $\mu$s (6.06 to 8.24 $\mu$s given the $T_1$ and Er concentration uncertainty); the similarity of this time to the fitted $T_2$ (7.5 $\mu$s) indicates that $T_2$ could be limited by spectral diffusion caused by $T_1$-induced flips of neighboring Er$^{3+}$ spins; in this case, the stretch factor $x$ in Eq. (2) would be 2. The resolution and signal to noise in the echo decay profile in Fig. 2(a) are insufficient to distinguish between $x = 1$ or $x = 2$; however, because of the indication that $T_2$ is limited by spectral diffusion, the fit was performed with $x$ fixed to 2. With $x$ fixed to 1, $T_2$ was calculated to be $10 \pm 3$ $\mu$s.

In summary, Er implanted Si is a promising platform for the development of QTs and is potentially highly scalable since it can utilize the silicon and ion implantation technology used in the IC industry. Er implanted Si can also exploit the atomic scale barrier to decoherence that is intrinsic to REs, and the recently developed ultra-low spin environment of isotopically purified $^{28}$Si, whereas the Er component itself is compatible with telecommunications wavelength photons and could be utilized for quantum communications schemes.

We report a spin coherence time of 7.5 $\mu$s at 5 K for $3 \times 10^{17}$ cm$^{-3}$ Er implanted Si; the spin-lattice relaxation time was $\sim 1$ ms. The calculated spectral diffusion time was 7.11 $\mu$s, which, because of its similarity to the measured $T_2$, indicated that $T_2$ was limited by spectral diffusion due to $T_1$-induced flips of neighboring Er$^{3+}$ spins. The origin of the echo is an Er center surrounded by six O atoms with monoclinic site symmetry. The spin echo decay profile had superimposed modulations due to strong superhyperfine coupling with a spin bath of $^{29}$Si nuclei.

See the supplementary material for the TRIM simulation of the implant profile and details of the background echo subtraction.

This work was supported by the UK EPSRC Grant Nos. EP/R011885/1 and EP/H026622/1. Authors acknowledge the European Research Council for financial support under the FP7 for the award of the ERC Advanced Investigator Grant No. SILAMPS 226470. We would like to thank Professor John Morton for helpful discussions.

The authors declare no competing financial interest.

## DATA AVAILABILITY

The data that support the findings of this study are openly available in Mendeley Data repository at http://dx.doi.org/10.17632/vt7vfm3jmb.1, Ref. 36.